\documentclass{ws-ijmpd}
\usepackage[super,compress]{cite}

\newcommand{\sfrac}[2]{{\textstyle{#1\over#2}}}
\newcommand{\N}[2]{\mathcal{N}^{#1}{}_{#2}}

\def\bom{\mbox{\boldmath $\omega$}}
\newcommand{\spur}{\mathop{\mathrm{tr}}}

\newcommand{\textfrac}[2]{{\textstyle{#1\over#2}}}

\newcommand{\lin}{l_{\mathrm{in}}}
\newcommand{\lout}{l_{\mathrm{out}}}

\renewcommand{\u}{\mathsf{u}}
\newcommand{\ue}{\check{u}}

\def\be{\begin{equation}}
\def\ee{\end{equation}}

\newcommand{\Thi}{\mathcal{T}_{\mathsf{Hi}}}

\newcommand{\Tco}{\mathcal{T}_{\mathsf{Jo}}}

\begin{document}

%
\catchline{}{}{}{}{}
%

\title{SPACETIME SINGULARITIES: 
RECENT DEVELOPMENTS\footnote{Based on a talk presented at Thirteenth Marcel Grossmann Meeting 
                             on General Relativity, Stockholm, July 2012.}}

\author{CLAES UGGLA}

\address{Department of Physics, University of Karlstad,\\
S-651 88 Karlstad, Sweden\\
Claes.Uggla@kau.se}

\maketitle


\begin{abstract}
Recent developments concerning oscillatory spacelike singularities in general
relativity are taking place on two fronts. The first treats generic
singularities in spatially homogeneous cosmology, most notably Bianchi types
VIII and IX. The second deals with generic oscillatory singularities in
inhomogeneous cosmologies, especially those with two commuting spacelike
Killing vectors. This paper describes recent progress in these two areas: in
the spatially homogeneous case focus is on mathematically rigorous results,
while analytical and numerical results concerning generic behavior and
so-called recurring spike formation are the main topic in the inhomogeneous
case. Unifying themes are connections between asymptotic behavior,
hierarchical structures, and solution generating techniques, which provide
hints for a link between the nature of generic singularities and a hierarchy
of hidden asymptotic symmetries.
\end{abstract}


\ccode{PACS numbers: 04.20.Dw, 04.20.Ha, 04.20.-q, 05.45.-a, 98.80.Jk}

\section{Introduction}\label{sec:intro}
This article describes and elaborates on the remarkable progress that has taken
place during the last few years regarding the nature of generic spacelike
singularities in general relativity. The developments have been taking place
on two fronts: (i) the nature of generic singularities in spatially
homogeneous (SH) Bianchi type VIII and IX models, and (ii) the nature of generic
spacelike singularities in inhomogeneous models, especially in models with
two commuting spacelike Killing vectors, the so-called $G_2$ models. To keep the
account reasonably short, considerations is restricted to 4-dimensional
vacuum spacetimes, which is a less restrictive assumption than one might
initially think. In their work, Belinski, Khalatnikov and Lifshitz
(BKL) provided heuristic evidence that some sources, like perfect fluids with
sufficiently soft equations of state such as dust or radiation, lead to
models that generically are asymptotically `\emph{vacuum dominated}', i.e.,
in the asymptotic appproach to a generic spacelike singularity, the spacetime geometry
is not influenced by the matter content, even though, e.g., the energy density
blows up~\cite{lk63,bkl70,bkl82}. However, arguably the most central, and
controversial, assumption of BKL is their \emph{`locality' conjecture}.
According to BKL, asymptotic dynamics toward a generic spacelike singularity
in \emph{inhomogeneous} cosmologies is \emph{`local},' in the sense that each
spatial point is assumed to evolve toward the singularity individually and
independently of its neighbors as a \emph{spatially homogeneous}
model.~\cite{lk63,bkl82}. It is no understatement to say that this conjecture
has set the stage for much of subsequent investigations about the detailed
nature of generic spacelike singularities.

A common unifying ingredient underlying recent progress at the two front
lines is the recasting of Einstein's field equations into \emph{scale
invariant asymptotically regularized dynamical systems} (first order systems
of autonomous ordinary differential equations (ODEs) and partial differential
equations (PDEs) in cases (i) and (ii), respectively) in the approach towards
a generic spacelike singularity. One of the advantages of such dynamical
systems, if appropriately defined, is that they lead to a \emph{state space
picture} with a \emph{hierarchy of invariant subsets, where simpler invariant
subsets constitute boundaries of more complex ones}. This hierarchical
structure is especially relevant in the context of generic spacelike
singularities since solutions of certain invariant `building block' subsets
at and near the `bottom' of a hierarchy can be joined into
`\emph{concatenated chains of solutions}' that describe the asymptotic
evolution along time lines toward the singularity. In the case of
inhomogeneous models this turns out to hold for `local BKL-like' behavior,
for which spatial derivatives can be neglected, as well as for `non-local
\emph{recurring spike behavior}' along certain time lines where spatial
derivatives cannot be neglected. Remarkably, all types of behavior are linked
to the solutions at the lowest invariant subset level of the hierarchy by
means of iterations of a \emph{solution generating algorithm}, which in turn
suggests that there exist asymptotic hidden symmetries, yet to be discovered.

The outline of the paper is as follows: section 2 focuses on recent
mathematically rigorous results concerning vacuum Bianchi type VIII and IX
spacetimes, while section 3 describes recent progress in the study of
generic singularities in inhomogeneous $G_2$ spacetimes. The concluding
section
relates the material in these two sections to the context of generic
singularities in general models without any symmetries. In particular
links are discussed between the nature of generic singularities and the hierarchy of
invariant subsets, which provide the building blocks for the asymptotic
construction of generic `temporally oscillatory' solutions toward generic
spacelike singularities.

\section{Bianchi type VIII and IX vacuum models}\label{sec:BIX}

The type VIII and~IX vacuum models belong to the `class~A' Bianchi vacuum
models for which the metric can be written as
\begin{subequations}\label{SHmetric}
\begin{align}
{}^4\mathbf{g} &= -dt\otimes
dt + g_{11}(t)\:\hat{\bom}^1\otimes \hat{\bom}^1 +
g_{22}(t)\:\hat{\bom}^2\otimes \hat{\bom}^2 +
g_{33}(t)\:\hat{\bom}^3\otimes \hat{\bom}^3, \label{threemetric}\\
d\hat{\bom}^1  &=  -\hat{n}_1 \, \hat{\bom}^2\wedge \hat{\bom}^3,\qquad
d\hat{\bom}^2  = -\hat{n}_2 \, \hat{\bom}^3\wedge \hat{\bom}^1,\qquad
d\hat{\bom}^3  =  -\hat{n}_3 \, \hat{\bom}^1\wedge \hat{\bom}^2,
\label{structconst}
\end{align}
\end{subequations}
where $\{\hat{\bom}^1,\hat{\bom}^2,\hat{\bom}^3\}$ is a symmetry-adapted
\mbox{(co-)}frame and where the constant parameters
$\hat{n}_1,\hat{n}_2,\hat{n}_3$ describe the structure constants of the
different class A Bianchi types (i.e. types I, II, VI$_0$, VII$_0$, VIII and
IX), see, e.g., Ref.~\citen{waiell97} and references therein.

Much of the recent mathematically rigorous progress is based on the scale
invariant `Hubble-normalized' dynamical system formulation of Einstein's
field equations for the SH `diagonal' class A Bianchi models, introduced by
Wainwright and Hsu~\cite{waihsu89} and generalized and elaborated on in the
book ``Dynamical Systems in Cosmology"~\cite{waiell97}.\,\footnote{The
dynamical systems approach and methods used in Ref.~\citen{waiell97} has many
precursors: notably work by Collins, Novikov, Bogoyavlensky, Rosquist,
Jantzen, Wainwright, CU, Coley, and many more, for references, see
Refs.~\citen{waiell97} and~\citen{col03}.} In this approach scale invariant
dimensionless variables are introduced by quotienting out the Hubble variable
$H$, which is related to the expansion $\theta$ of the normal congruence of
the SH symmetry surfaces and $\spur k$, the trace of the second fundamental
form $k_{\alpha\beta}$, according to $H=\theta/3 = -\spur k/3$. This yields
the dimensionless variables\footnote{In Bianchi types~VIII and IX there is a
one-to-one correspondence between $g_{11}, g_{22}, g_{33} $ and $N_1, N_2,
N_3$. For the lower Bianchi types~I--$\mathrm{VII}_0$, some of the variables
$N_1,N_2,N_3$ are zero, cf.~\eqref{Hnormvars}; in this case, the other
variables, i.e., $H, \Sigma_\alpha$, are needed as well to reconstruct the
metric; see Ref.~\citen{janugg99} for a group theoretical approach.}
\begin{subequations}\label{Hnormvars}
\begin{alignat}{5}
\Sigma_1 &:= \frac{k^1\!_1 - \sfrac13\spur k}{\sfrac13\spur k},&\quad
\Sigma_2 &:= \frac{k^2\!_2 - \sfrac13\spur k}{\sfrac13\spur k},&\quad
\Sigma_3 &:= \frac{k^3\!_3 - \sfrac13\spur k}{\sfrac13\spur k},\\
N_1 &:= -\frac{\hat{n}_1}{\sfrac13\spur
k}\,\sqrt{\frac{g_{11}}{g_{22}g_{33}}},&\quad N_2 &:=
-\frac{\hat{n}_2}{\sfrac13\spur k}\,\sqrt{\frac{g_{22}}{g_{33}g_{11}}},&\quad
N_3 &:= -\frac{\hat{n}_3}{\sfrac13\spur
k}\,\sqrt{\frac{g_{33}}{g_{11}g_{22}}},
\end{alignat}
\end{subequations}
and hence $\Sigma_1 + \Sigma_2 + \Sigma_3 = 0$ ($k^1\!_1 = -\frac12
g_{11}^{-1}{dg_{11}}/{dt}$, and similarly for $k^2\!_2$ and
$k^3\!_3$).\footnote{It is common to globally solve $\Sigma_1 + \Sigma_2 +
\Sigma_3 = 0$ by introducing new variables according to $\Sigma_1 = -2
\Sigma_+$, $\Sigma_2 = \Sigma_+ - \sqrt{3} \Sigma_-$, $\Sigma_3 = \Sigma_+ +
\sqrt{3} \Sigma_-$, or a permutation thereof, which yields $\Sigma^2 =
\Sigma_+^2 + \Sigma_-^2$, but this unfortunately breaks the permutation
symmetry in Bianchi type~IX.} In addition a new dimensionless time variable
$\tau$ is defined according to $d\tau/dt=H$.

When the (vacuum) Einstein  field equations are reformulated in terms of the
dimensional Hubble variable $H$ and the dimensionless `Hubble-normalized'
variables $(\Sigma_\alpha, N_\alpha)$ it follows from dimensional reasons
that the equation for $H$,
\begin{equation}\label{H}
H^\prime = -(1 + 2\Sigma^2)H, \qquad \Sigma^2 := \sfrac{1}{6}(\Sigma_1^2 + \Sigma_2^2 + \Sigma_3^2),
\end{equation}
decouples from the remaining equations, which form   the following coupled
system of ODEs:~\cite{heiugg09b}
\begin{subequations}\label{IXeq}
\begin{align}
\label{sig}
\Sigma_\alpha^\prime & =  -2(1-\Sigma^2)\Sigma_\alpha - \sfrac{1}{3}\!\left[ N_\alpha(2N_\alpha - N_\beta
- N_\gamma) - (N_\beta - N_\gamma)^2 \right], \\[0.5ex]
\label{n}
N_\alpha^\prime & =  2(\Sigma^2 + \Sigma_\alpha)\,N_\alpha
\qquad\qquad\qquad \text{(no sum over $\alpha$)},\\
\label{gauss}
1 &= \Sigma^2 +
\sfrac{1}{12} \Big[ N_1^2 + N_2^2 + N_ 3^2 -
2 \left( N_1N_2 + N_2 N_3 + N_3N_1 \right)\Big],
\end{align}
\end{subequations}
where $(\alpha\beta\gamma) \in \left\{(123),(231),(312)\right \}$
in~\eqref{sig}, and where a prime denotes the derivative $d/d\tau$.

It follows from the Gauss constraint~\eqref{gauss} that $H$ remains positive
if $H$ is positive initially for all vacuum class A models except type~IX. In
Bianchi type~IX, however, a theorem by Lin and Wald~\cite{linwal89} implies
that all vacuum models first expand ($H>0$), reach a point of maximum
expansion ($H=0$), and then re-collapse ($H<0$). In this case the variables
$(\Sigma_\alpha, N_\alpha)$ break down at the point of maximum expansion, but
they give a correct description of the dynamics in the expanding phase which
we will focus on henceforth.\footnote{For bounded variables that do not break
down at the point of maximum expansion and allows for a global description
of the dynamics, see Ref.~\citen{heiugg09a}.}

Since $\Sigma_1 + \Sigma_2 + \Sigma_3 = 0$, equations~\eqref{Hnormvars}
and~\eqref{IXeq} imply that the \emph{dimensionless state space} of the
Bianchi type IX and VIII vacuum models is 4-dimensional (all $N_\alpha$ are
non-zero and have the same sign in type IX, while in type VIII one $N_\alpha$
has an opposite sign compared to the others, which is due to the signs of
$\hat{n}_\alpha$, which determine the symmetry group type). Each lower
Bianchi type than types IX and VIII in class A can be obtained by means of
Lie contractions, i.e., by setting one, two, and eventually all three of the
constants $\hat{n}_1,\hat{n}_2,\hat{n}_3$ to zero. The associated
dimensionless state spaces are described by setting the corresponding
$N_\alpha$ to zero (recall that $N_\alpha \propto \hat{n}_\alpha$), which
corresponds to invariant boundary subsets of~\eqref{IXeq}. The
system~\eqref{IXeq} thus exhibits a hierarchical invariant boundary subset
structure: In Bianchi types VII$_0$ and VI$_0$ two of the variables
$N_\alpha$ are non-zero, and the dimensionless state space is therefore
3-dimensional; in Bianchi type II only one $N_\alpha$ is non-zero, which
leads to a 2-dimensional state space, while in type I all $N_\alpha$ are
zero, which hence results in a 1-dimensional state space.

The hierarchical `Lie contraction boundary subset structure' of~\eqref{IXeq}
is central for the asymptotic dynamics toward the initial singularity. Every
time a constant $\hat{n}_\alpha$ is set to zero the dimension of the
automorphism group increases by one (in this context automorphisms are
linear constant transformations of the symmetry adapted spatial frame
$\{\hat{\bom}^1,\hat{\bom}^2,\hat{\bom}^3\}$ that leave the structure
constants unchanged). The \emph{kinematical} consequence of this is that a
given Lie contracted boundary subset of~\eqref{IXeq} describes the true
degrees of freedom of the associated Bianchi type. Due to this, the metric of
that Bianchi type can be explicitly constructed from the solution to the
equations on the subset, and the quadrature for the decoupled variable $H$ by
integrating~\eqref{H}, whose decoupling is a consequence of the scale
invariance symmetry, see Ref.~\citen{janugg99}.

More importantly, however, are the \emph{dynamical} implications of the group
of automorphisms and scale transformations. As explicitly shown in
Ref.~\citen{heiugg10}, on each level in the `\emph{Lie contraction boundary
subset hierarchy´}' the combined scale-automorphism group induces monotone
functions, and even constants of the motion at the bottom of the hierarchy.
The resulting \emph{hierarchy of monotone functions} pushes the dynamics
towards the past singularity to boundaries of boundaries in the hierarchy,
where the solutions on the simplest subsets, i.e., those of Bianchi types I
and II, are completely determined by scale and automorphism symmetries. The
dynamics towards the initial singularity hence turns out to be governed to a
large extent by structures induced by the scale-automorphism groups on the
different levels in the Lie contraction boundary subset hierarchy. Since the
automorphisms in the present context correspond to the spatial diffeomorphism
freedom that respects the symmetries of the various Bianchi models, it
therefore follows that the dynamics towards the initial singularity is partly
determined by physical first principles, namely scale invariance and general
covariance.\footnote{In the case of matter sources, hierarchies become even
more important than in the vacuum case. Then, in addition to Lie
contractions, one also have source contractions, where the vacuum is at the
bottom of the source hierarchy. For each level of the source contraction
hierarchy the scale-automorphism group yields different structures, such as
monotone functions, leading to restrictions on asymptotic dynamics; see
Ref.~\citen{heiugg10} where this general feature is exemplified explicitly by
a perfect fluid in the case of diagonal class A Bianchi models.}

As a consequence of the hierarchical structure, it is both natural and
necessary to describe the dynamics from the bottom up, i.e., from Bianchi type I
and upwards in the Lie contraction hierarchy. This is conveniently done by
projecting the dynamics onto $(\Sigma_1, \Sigma_2,\Sigma_3)$ space.

\begin{figure}[ht]
\centering
        \includegraphics[width=0.55\textwidth]{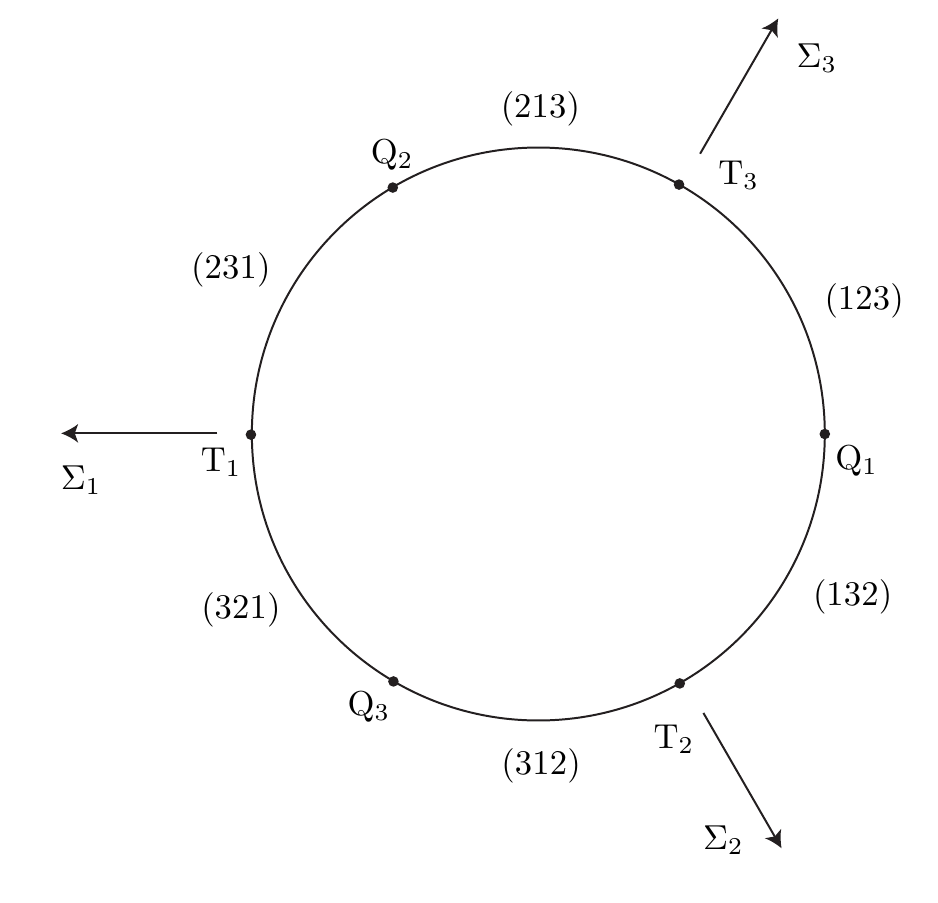}
\caption{The division of the Kasner circle $\mathrm{K}^\mathrm{O}$ of fixed points
into six equivalent sectors and six LRS fixed points
$\mathrm{T}_\alpha$ and $\mathrm{Q}_\alpha$, $\alpha = 1,2,3$. Sector $(\alpha\beta\gamma)$ is defined by
$\Sigma_\alpha < \Sigma_\beta < \Sigma_\gamma$.}\label{fig:plensectors}
\end{figure}

One of the advantages of scale invariant approaches such as the
Hubble-normalized approach is that \emph{self-similar}, i.e., scale
invariant, models appear as fixed points in the dimensionless state
space~\cite{rosjan85,janros86,hsuwai86}. This is the case for the Bianchi
type I vacuum models, the `Kasner' solutions, which in the Hubble-normalized
state space picture form the so-called \textit{Kasner circle}
$\mathrm{K}^{\rm O}$ of fixed points, characterized by $N_1 = N_2 = N_3 = 0$,
$\Sigma^2 =1$, and constant $\Sigma_1, \Sigma_2, \Sigma_3$. Although points
on $\mathrm{K}^{\rm O}$ are characterized by $\Sigma_1,\Sigma_2,\Sigma_3$,
the Kasner solutions are often characterized by the constant Kasner exponents
$p_1,p_2,p_3$, $p_1+p_2+p_3=p_1^2+p_2^2+p_3^2$,\footnote{The 1-parameter
family of Kasner solutions is often given in terms of the line element $ds^2
= -dt^2 + t^{2p_1}dx^2 + t^{2p_2}dy^2 + t^{2p_3}dz^2$.} where the Kasner
exponents are related to the $\Sigma_\alpha$ variables via the relation
$\Sigma_\alpha=3p_\alpha- 1$. Due to permutations of the axes,
$\mathrm{K}^{\rm O}$ is naturally divided into six equivalent sectors,
denoted by permutations of the triple $(123)$ where sector
$(\alpha\beta\gamma)$ is defined by $\Sigma_\alpha < \Sigma_\beta <
\Sigma_\gamma$, or, equivalently, $p_\alpha < p_\beta < p_\gamma$, see
Fig.~\ref{fig:plensectors}. The boundaries of the sectors are six special
points that are associated with locally rotationally symmetric (LRS)
solutions,
\begin{subequations}
\begin{alignat}{5}
\mathrm{T}_\alpha\!: \,\, & (\Sigma_\alpha, \Sigma_\beta, \Sigma_\gamma)\,
&=&\, ({+2},{-1},{-1}), \,\, &\text{or, equivalently,} \,\,
&(p_\alpha,p_\beta,p_\gamma)\,
&=&\, (1,0,0), \\
\mathrm{Q}_\alpha\!: \,\, & (\Sigma_\alpha, \Sigma_\beta, \Sigma_\gamma)\,
&=&\, ({-2},{+1},{+1}), \,\, &\text{or, equivalently,} \,\,
&(p_\alpha,p_\beta,p_\gamma)\,
&=&\,(-\textfrac{1}{3},\textfrac{2}{3},\textfrac{2}{3}).
\end{alignat}
\end{subequations}
The \textit{Taub points} $\mathrm{T}_\alpha$ ($\alpha = 1,2,3$) correspond to
the Taub (LRS) representations of the Minkowski spacetime, while
$\mathrm{Q}_\alpha$ yield three equivalent LRS solutions with non-flat
geometry.

%

%
%
%
%

It is useful to parameterize $\mathrm{K}^{\rm O}$ with \emph{extended Kasner
parameters}~\cite{damlec11a,heietal12}. To each cyclic permutation of $(123)$
there exists a natural parameter, $\ue_\alpha \in (-\infty,\infty)$, defined
by
\begin{equation}\label{ueeq}
p_\alpha = -\ue_\alpha/f(\ue_\alpha),\qquad
p_\beta = (1+\ue_\alpha)/f(\ue_\alpha),\qquad
p_\gamma = \ue_\alpha(1+\ue_\alpha)/f(\ue_\alpha),
\end{equation}
where $(\alpha\beta\gamma) = (123)$ and cycle, and where
\begin{equation}
f = f(x) := 1 + x + x^2.
\end{equation}
Factoring out the permutation freedom leads to the usual  \emph{frame (gauge)
invariant} Kasner parameter $u\in(1,\infty)$ (note the difference in
interval compared to that for $\ue_\alpha$), which is defined according
to~\cite{bkl70,khaetal85}
\begin{equation}\label{ueq}
p_\alpha = - u/f(u)\,,\qquad p_\beta = (1+u)/f(u)\,,\qquad
p_\gamma = u(1+u)/f(u),
\end{equation}
where the boundary points of sector $(\alpha\beta\gamma)$,
$\mathrm{Q}_\alpha$ and $\mathrm{T}_\gamma$, are characterized by $u=1$ and
$u=\infty$, respectively.\footnote{The Kasner parameter $u$ can be related to
a function that is constructed entirely from Weyl scalars in a one-to-one
manner, and hence $u$ is a gauge invariant quantity.}

Comparing~\eqref{ueq} and~\eqref{ueeq} gives a transformation between $u$ and
$\ue_\alpha$ for each sector:
\begin{subequations}\label{ueandu}
\begin{xalignat}{2}
& (\alpha\beta\gamma):\, (1,\infty) \ni \ue_\alpha = u, \quad
& & (\alpha\gamma\beta):\, (0,1) \ni \ue_\alpha = u^{-1},  \\
&  (\gamma\alpha\beta):\, (-\textfrac{1}{2}, 0) \ni \ue_\alpha
=-\frac{1}{u+1},\quad
& & (\gamma\beta\alpha):\, (-1,-\textfrac{1}{2}) \ni \ue_\alpha = -\frac{u}{u+1}, \\
&  (\beta\gamma\alpha):\, (-2,-1) \ni \ue_\alpha =-\frac{u+1}{u},\quad & &
(\beta\alpha\gamma):\, (-\infty,-2) \ni \ue_\alpha = -(u+1),
\end{xalignat}
\end{subequations}
where $(\alpha\beta\gamma) = (123)$ and cycle, which yields
%
%
%
\begin{subequations}\label{checkutransf}
\begin{alignat}{3}
\check{u}_1 &= -\frac{\check{u}_2 + 1}{\check{u}_2}, &\qquad \check{u}_2 &=
-\frac{\check{u}_3 + 1}{\check{u}_3},&\qquad \check{u}_3 &=
-\frac{\check{u}_1 + 1}{\check{u}_1},\\
\check{u}_1 &= -\frac{1}{\check{u}_3 + 1}, &\qquad \check{u}_2 &=
-\frac{1}{\check{u}_1 + 1},&\qquad \check{u}_3 &= -\frac{1}{\check{u}_2 + 1}.
\end{alignat}
\end{subequations}

\begin{figure}[th]
\centering
\includegraphics[width=0.30\textwidth]{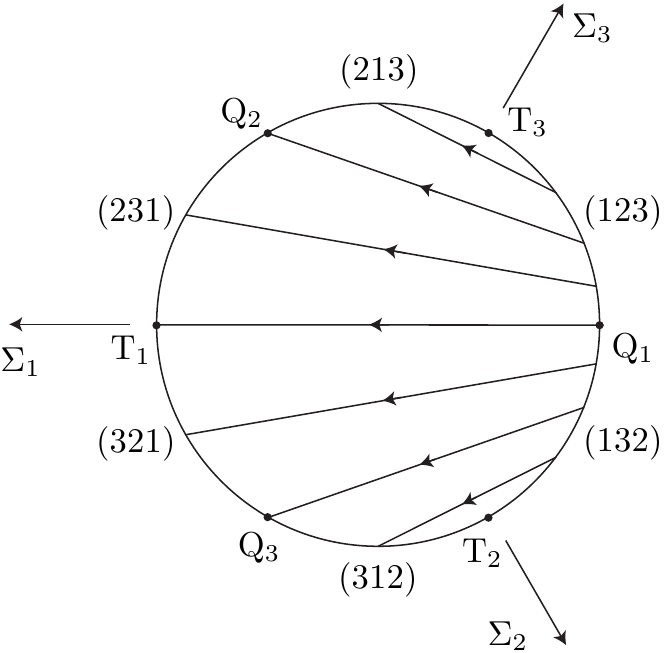}\,
\includegraphics[width=0.30\textwidth]{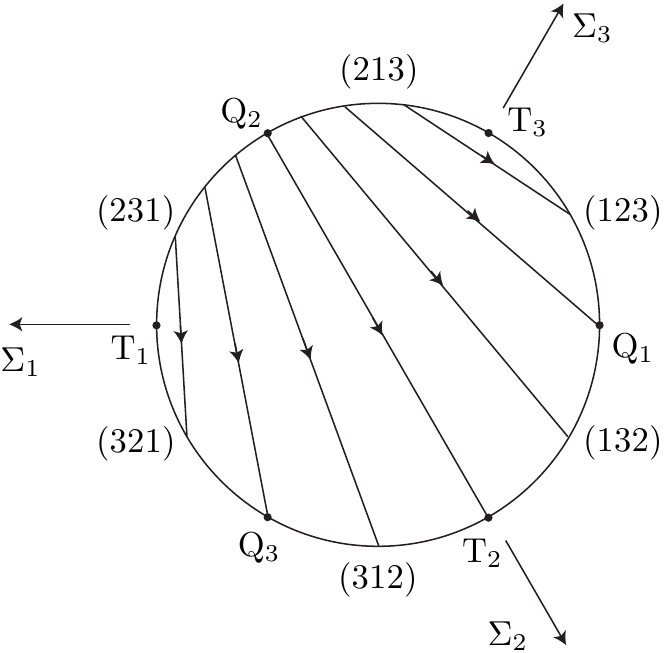}\,
\includegraphics[width=0.30\textwidth]{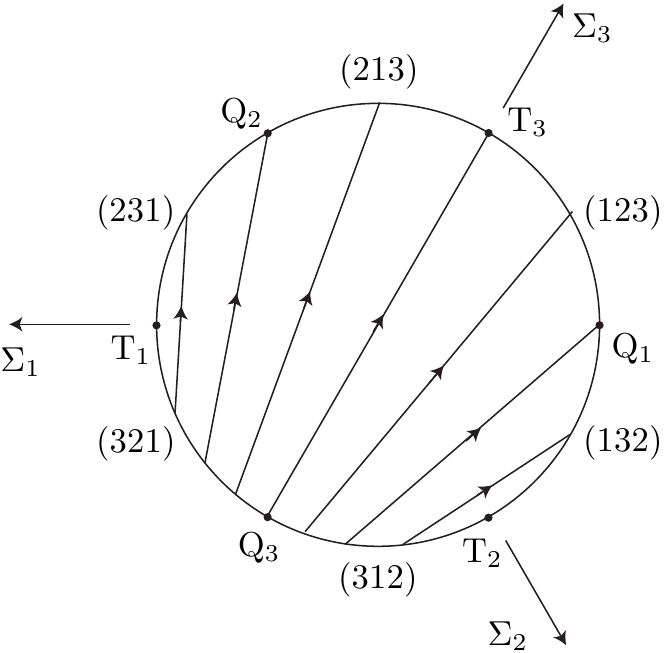}
\caption{Projections of Bianchi type~II transitions
$\mathcal{T}_{N_1}$, $\mathcal{T}_{N_2}$, $\mathcal{T}_{N_3}$
on the Bianchi type~II subsets $\mathcal{B}_{N_1}$, $\mathcal{B}_{N_2}$, $\mathcal{B}_{N_3}$ (from left
to right in the figure) onto $(\Sigma_1, \Sigma_2,\Sigma_3)$ space.
The arrows indicate the direction of time towards the past.}
\label{Fig:alltypeII}
\end{figure}

The next level in the Lie contraction hierarchy consists of the Bianchi type
II subsets $\mathcal{B}_{N_\alpha}$ given by $N_\alpha \neq 0$, $N_\beta =
N_\gamma = 0$. The solutions of these subsets are also, as in the Bianchi
type I case, completely determined by the scale-automorphism group. Projected
onto $(\Sigma_1,\Sigma_2,\Sigma_3)$ space they form families of straight
lines (see e.g. Refs.~\citen{heietal09} and~\citen{heiugg09b}) where each
straight line connects two fixed points on $\mathrm{K}^\mathrm{O}$, see
Fig.~\ref{Fig:alltypeII}.

It follows that each solution trajectory is a so-called \emph{heteroclinic
orbit}~\cite{waiell97}, i.e., a solution trajectory that starts
($\alpha$-limit) and ends ($\omega$-limit) at two different fixed points. In
Refs.~\citen{heiugg09b} and~\citen{heietal09} these orbits were denoted as
Bianchi type~II \textit{transitions}, $\mathcal{T}_{N_\alpha}$
($\alpha=1,2,3$), because each orbit can be viewed as representing a
transition from one Kasner state to another. It follows that the type II
vacuum models are past and future \emph{asymptotically self-similar}, since
all orbits begin and end at two different Kasner fix points that correspond
to self-similar Kasner solutions.

In the context of Bianchi types VIII and IX, it follows from the properties
of a $\mathcal{T}_{N_\alpha}$ transition on the $\mathcal{B}_{N_\alpha}$
boundary that it gives rise to a `\emph{Mixmaster map}' between two fixed
points on $\mathrm{K}^{\rm O}$ in the \emph{direction towards the past}
initial singularity according to:~\cite{heietal12}
\begin{equation}\label{mixmastermap}
\mathcal{T}_{N_\alpha}\!\!: \,\,\, \ue_\alpha = \ue_\alpha^{\mathrm{i}} \in (0,+\infty)\quad \mapsto \quad
\ue_\alpha^{\mathrm{f}} = -\ue_\alpha^{\mathrm{i}} \in (-\infty,0).
\end{equation}
Quoting out the gauge dependence, the Mixmaster map yields the \emph{Kasner
map} (also known as the BKL map):~\cite{bkl70,khaetal85}
\begin{equation}\label{BKLMap}
u^{\mathrm{f}}  \:= \: \left\{\begin{array}{ll}
u^{\mathrm{i}} - 1 & \qquad \text{if}\quad u^{\mathrm{i}} \in [2, \infty), \\[1ex]
(u^{\mathrm{i}} - 1)^{-1} &\qquad  \text{if} \quad u^{\mathrm{i}} \in [1,2].
\end{array}\right.
\end{equation}

Based on work reviewed and developed in Ref.~\citen{waiell97} and by
Rendall~\cite{ren97}, Ringstr\"om, in 2000 and 2001, produced the first major
proofs about asymptotic Bianchi type~IX dynamics~\cite{rin00,rin01}. In
particular Ringstr\"om managed to prove that the past attractor in Bianchi
type IX \emph{resides} on a subset that consists of the union of the Bianchi
type~I and~II vacuum subsets (for a shorter proof that uses the complete
structure induced by the scale-automorphism group, see
Ref.~\citen{heiugg09a}), i.e.,
\begin{equation}\label{AIXdef}
\mathcal{A} = \mathrm{K}^\mathrm{O} \cup
\mathcal{B}_{N_1} \cup
\mathcal{B}_{N_2} \cup
\mathcal{B}_{N_3}.
\end{equation}
This achievement is impressive, but Ringstr\"om's `attractor theorem' does
not say if all of $\mathcal{A}$ is the attractor nor if the Kasner map is
relevant for dynamics asymptotic to the initial singularity in Bianchi type
IX, and the theorem says nothing about Bianchi type VIII; for further
discussion, see Ref.~\citen{heiugg09b}. The Kasner map~\eqref{BKLMap}, when
iterated, turns out to be associated with chaotic properties that has
attracted considerable attention, see Ref.~\citen{heiugg09b} for references.
Taken together with BKL's locality conjecture, these properties are often
said to imply that generic spacelike singularities, and hence also Einstein's
equations, are chaotic. But up until recently there were no rigorous results,
including Ringstr\"om's theorems, that in any way tied the Mixmaster and
Kasner maps to asymptotic dynamics. For example, there was nothing that
excluded that
some periodic sequence(s) of orbits (further discussed
below), associated with a particular value (values) of $u$, could not be the past
attractor, which then would lead to a simple analytic asymptotic description
that in no way could be called chaotic. Nor is it possible that any numerical
experiment can shed any light on this since it follows from continuity, and
the transversal hyperbolicity of $\mathrm{K}^{\mathrm{O}}$, that there are
solution trajectories that shadow orbits associated with any sequence of $u$
obtained by iterations of~\eqref{BKLMap} for arbitrarily long \emph{finite}
$\tau$ intervals. Fortunately, during the last couple of years there has been
substantial progress that has begun to rigorously asymptotically connect the
structures on ${\cal A}$ with the Mixmaster and Kasner maps.

\begin{figure}[ht]
 \centering{
  \includegraphics[height=0.45\textwidth]{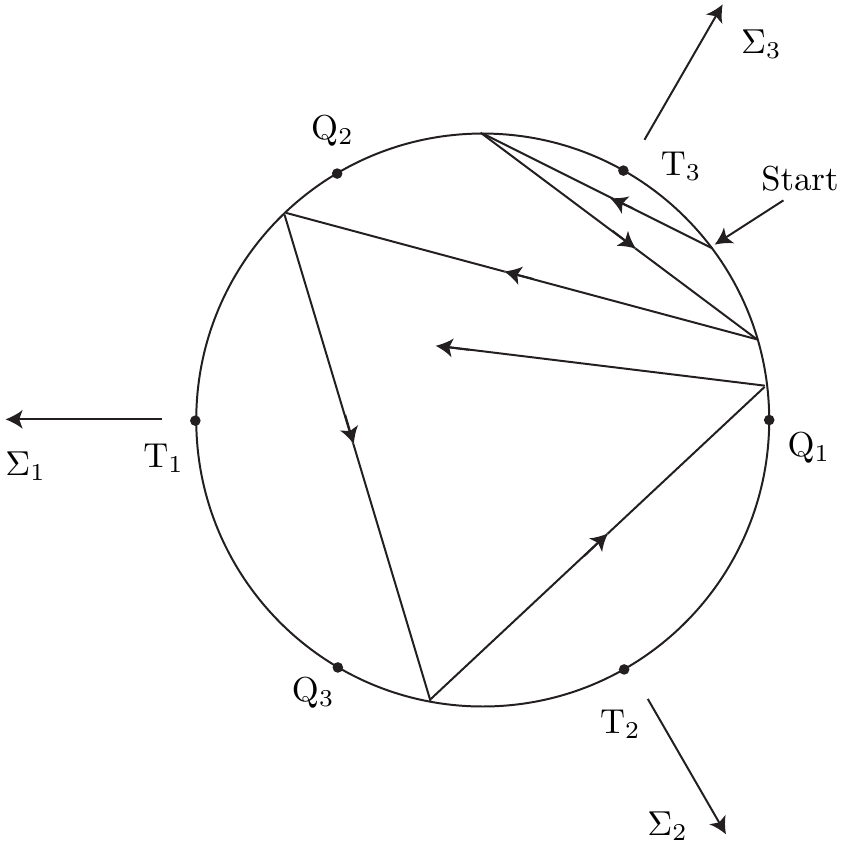}}
\caption{Concatenating Bianchi type~II transition orbits
on the Bianchi type II and I boundaries in Bianchi types VIII and IX
leads to sequences of transitions---heteroclinic Mixmaster chains.
The discrete map governing the associated sequence of fixed points on
$\mathrm{K}^\mathrm{O}$ is the Mixmaster map, which, when axis permutations
are quotiented out, yields the Kasner map. The arrows indicate the
direction of time towards the past.}
\label{Fig:Mixmaster}
\end{figure}

To understand these new results we need to focus on the \emph{heteroclinic
structure} associated with $\mathcal{A}$. The heteroclinic orbits
$\mathcal{T}_{N_1}, \mathcal{T}_{N_2}, \mathcal{T}_{N_3}$ can be
\emph{concatenated} on $\mathcal{A}$ to yield heteroclinic \emph{`Mixmaster'
chains} by identifying the `final' fixed point ($\omega$-limit point) of one
transition with the `initial' fixed point ($\alpha$-limit point) of another
transition (in dynamical systems theory a heteroclinic chain is defined as a
sequence of heteroclinic orbits such that the $\omega$-limit point of one
orbit is the $\alpha$-limit point of the subsequent orbit), see
Fig.~\ref{Fig:Mixmaster}.

The Mixmaster chains induce iterations of the maps~\eqref{mixmastermap}
and~\eqref{BKLMap}. The iterations of the Mixmaster map can be analytically
described by combining~\eqref{mixmastermap} with~\eqref{checkutransf}, which
allows one to describe the sequence of Kasner fixed points obtained via
Bianchi type II transitions by means of a single extended Kasner parameter,
e.g. $\ue_1$. In terms of the Kasner parameter $u$, a sequence of transitions
corresponds to an iteration of~\eqref{BKLMap}. Let $l=0,1,2,\ldots$ and let
$u_l$ denote the initial Kasner state of the $l$\raisebox{0.7ex}{\small th}
transition, then the iterated Kasner map is given by:
\begin{equation}\label{Kasnermap}
u_l \:\,\xrightarrow{\;\text{$l$\raisebox{0.5ex}{th} transition}\;}\:\, u_{l+1}:
\qquad\quad
u_{l+1} \:= \: \left\{\begin{array}{ll}
u_l - 1 & \qquad \text{if}\quad u_l \in[2,\infty), \\[1ex]
(u_l - 1)^{-1} &\qquad  \text{if} \quad u_l \in [1,2].
\end{array}\right.
\end{equation}
Since each value of the Kasner parameter $u \in(1,\infty)$ represents an
equivalence class of six Kasner fixed points, the Kasner map can be regarded
as the map induced by the Mixmaster map on these equivalence classes via the
equivalence relation.

In a sequence $(u_l)_{l=0,1,2,\ldots}$ that is generated by the Kasner
map~\eqref{Kasnermap}, each Kasner state $u_l$ is called a \textit{Kasner
epoch}. Every sequence $(u_l)_{l=0,1,2,\ldots}$ possesses a natural partition
into pieces called \emph{Kasner era}s with a finite number of epochs. An era
begins with a maximal value $u_{\lin}$ (where $u_{\lin}$ is generated from
$u_{\lin-1}$ by $u_{\lin} = [u_{\lin-1}-1]^{-1}$), and continues with a
sequence of Kasner parameters obtained via $u_l \mapsto u_{l+1} = u_l - 1$;
it ends with a minimal value $u_{\lout}$ that satisfies $1 < u_{\lout} < 2$,
so that $u_{\lout+1} = [u_{\lout}-1]^{-1}$ begins a new era~\cite{bkl70}, as
exemplified by
\begin{equation}\label{phases}
\underbrace{3.45 \rightarrow 2.45 \rightarrow 1.45}_{\text{\scriptsize era}}  \rightarrow
\underbrace{2.23 \rightarrow 1.23}_{\text{\scriptsize era}}  \rightarrow
\underbrace{4.33 \rightarrow 3.33 \rightarrow 2.33 \rightarrow \ldots}_{\text{\scriptsize era}}
\end{equation}

Denoting the initial and maximal value of the Kasner parameter $u$ in era
number $s$ (where $s = 0,1,2,\ldots$) by $\u_s$,  and decomposing $\u_s$ into
its integer $k_s = [\u_s]$ and fractional $x_s = \{\u_s\}$ parts,
gives~\cite{bkl70,khaetal85}
\begin{equation}\label{usdecomp}
\u_s = k_s + x_s,
\end{equation}
where $k_s$ represents the (discrete) length of era $s$, and hence the number
of Kasner epochs it contains. The final and minimal value of the Kasner
parameter in era $s$ is given by $1 + x_s$, which implies that era number
$(s+1)$ begins with
\begin{equation}
\u_{s+1} = \frac{1}{x_s} = \frac{1}{\{\u_s\}}.
\end{equation}
The map $\u_s \mapsto \u_{s+1}$ is the so-called `\emph{era map}.' Starting
from $\u_0 = u_0$ it recursively determines $\u_s$, $s=0,1,2,\ldots$, and
thereby the complete Kasner sequence $(u_l)_{l=0,1,\ldots}$.

The era map admits an interpretation in terms of continued fractions.
Applying the Kasner map to the continued fraction representation of the
initial value $\u_0$,
\begin{equation}
\u_0 =  k_0 + \cfrac{1}{k_1 + \cfrac{1}{k_2 + \dotsb}} = [k_0; k_1,k_2,k_3,\dotsc]\,,
\end{equation}
gives
\begin{align}
\begin{split}
u_0 = \mathsf{u}_0 & =
\big[ k_0; k_1, k_2,  \dotsc \big] \rightarrow  \big[ k_0 -1 ; k_1, k_2 , \dotsc \big]
\rightarrow \ldots \rightarrow \big[1 ; k_1, k_2 , \dotsc \big] \\
 \rightarrow \mathsf{u}_1 & = \big[ k_1; k_2 , k_3, \dotsc \big]
 \rightarrow  \big[  k_1-1; k_2 , k_3, \dotsc \big]
\rightarrow \ldots \rightarrow \big[1 ; k_2, k_3 , \dotsc \big] \\
 \rightarrow \mathsf{u}_2 & = \big[ k_2; k_3 , k_4, \dotsc \big]
 \rightarrow  \big[  k_2-1; k_3 , k_4, \dotsc \big]
\rightarrow \ldots\,,
\end{split}
\end{align}
and hence the era map is simply a shift to the left in the continued fraction
expansion,
\begin{equation}
  \u_s = [k_s; k_{s+1}, k_{s+2}, \dotsc] \:\mapsto\:
  \u_{s+1} = [k_{s+1}; k_{s+2}, k_{s+3},\dotsc]\:.
\end{equation}

Some of the \emph{era} and \emph{Kasner sequences\/} are periodic, notably
$u_0 = [(1)] = [1;1,1,1,\dotsc]= (1+\sqrt{5})/2$, which is the golden ratio,
gives $\u_s = (1+\sqrt{5})/2$ $\forall s$, and hence the Kasner sequence is
also a sequence with period $1$,
\begin{equation*}
(u_l)_{l\in\mathbb{N}}:\quad \sfrac{1}{2}\big(1+\sqrt{5}\big) \rightarrow
\sfrac{1}{2}\big(1+\sqrt{5}\big) \rightarrow
\sfrac{1}{2}\big(1+\sqrt{5}\big) \rightarrow
\sfrac{1}{2}\big(1+\sqrt{5}\big) \rightarrow
\ldots ,
\end{equation*}
while this yields two heteroclinic cycles of period 3 in the state space
picture (since the axis permutations are not quotiented out in the state
space), see the figures in Ref.~\citen{heiugg09b}.

Although BKL~\cite{bkl70,bkl82,khaetal85}, as well as
Misner~\cite{mis69a,mis69b}, conjectured that the asymptotic dynamics of
Bianchi type IX is governed by the Kasner and era maps, it was only recently that
rigorous results were obtained relating these maps to asymptotic dynamics in
Bianchi types VIII and IX. To describe these results, it is convenient to use
a modified version of the classification scheme of Kasner sequences and the
associated Mixmaster chains that was introduced in Ref.~\citen{heiugg09b}:
%
\begin{itemize}
\item[(i)]$u_0 = [k_0; k_1, k_2, \dotsc, k_n ]$, i.e., $u_0
    \in\mathbb{Q}$. The associated Kasner sequence is finite with $n$
    eras and have an associated Mixmaster chain that terminates at one of
    the Taub points. It has been proven that these sequences are not
    asymptotically realized in the generic non-LRS case since a Taub
    point is not the $\omega$-limit set of any non-LRS
    solution~\cite{rin00,rin01,heiugg09a}.
\item[(ii)] $u_0 = [k_0;k_1,\dotsc]$ such that the sequence of partial
    quotients of its continued fraction representation is bounded, with
    or without periodicity. As a consequence the associated Mixmaster
    chains avoid a (small) neighborhood of the Taub points. In the case
    of no periodicity and no cycles, B\'eguin proved that a family of
    solutions of codimension one converges to each associated
    chain~\cite{beg10} (for the proof to work, cycles must be excluded to
    avoid resonances). By using different techniques, and different
    differentiability conditions, Liebscher {\it et
    al\/}~\cite{lieetal10} proved explicitly that a family of solutions
    of codimension one converges to the $3$-cycle associated with $u_0 =
    [(1)]$. The authors also gave arguments about how their methods could
    be extended to the present general case. This was explicitly proved
    in Ref.~\citen{lieetal12}, where the authors introduced a new
    technique that involves the invariant Bianchi type I and II subset
    structure. This is a quite promising development since this structure
    is related to the Lie contraction hierarchy, which in turn is tied to
    basic physical principles that characterize the problem at hand.
\item[(iii)] $u_0 = [k_0;k_1,\dotsc]$ is an unbounded sequence of partial
    quotients, which is the generic case. The associated Kasner sequence
    is unbounded and the associated Mixmaster chain enters every
    neighborhood of the Taub points infinitely often. As argued in
    Ref.~\citen{reitru10},\footnote{Ref.~\citen{reitru10} uses quite
    different mathematical techniques than the other rigorous papers in
    this area. As a consequence the results seem to be somewhat
    controversial in the research community, although the claims are
    arguably quite plausible. Due to this, and due to an intrinsic value,
    it would be of interest if the results could be confirmed, or
    preferably even extended, with some other independent methods.} a
    subset of these chains is relevant to the description of the
    asymptotic dynamics of actual solutions: for each $u_0$ such that the
    sequence $(k_n)_{n\in\mathbb{N}}$ can be bounded by a function of $n$
    with a prescribed growth rate, there exists an actual solution that
    converges to the chain determined by $u_0$. On the other hand, chains
    associated with initial values $u_0 = [k_0;k_1,\dotsc]$ with rapidly
    increasing partial quotients $k_n$, $n\in\mathbb{N}$ are perhaps less
    relevant for the description of the asymptotic dynamics of actual
    solutions; if a solution shadows a finite part of such a chain it may
    be thrown off course at the point where the chain enters a (too)
    small neighborhood of the Taub points. The prescribed growth rate
    given in Ref.~\citen{reitru10} is weak enough to not destroy
    genericity; a generic real number has a continued fraction
    representation compatible with the required boundedness condition.
    Note, however, that these results do not say anything about how many
    solutions actually converge to a given chain, nor if the
    asymptotic dynamics of a generic initial data set is represented by a
    heteroclinic chain.
\end{itemize}

The above results imply that $\mathcal{A}$ is indeed the global past
attractor for Bianchi type IX, but $\mathcal{A}$ is not necessarily the
global past attractor for type VIII, since it still has not been excluded
that the type VIII attractor also involves the vacuum Bianchi type VI$_0$
subset. It is also worth mentioning that the general Bianchi type VI$_{-1/9}$
models are as general as those of types VIII and IX, but arguably they are
more relevant for generic singularities~\cite{heietal09}. They also have an
oscillatory singularity, and hence asymptotic `\emph{self-similarity
breaking}'~\cite{waietal99}, but instead of being characterized by Mixmaster
chains, the singularity is conjectured to be characterized by so-called
Iwasawa chains, see Ref.~\citen{heietal09}. Unfortunately, there exist no
rigorous mathematical results concerning their past asymptotic dynamics.

\section{Inhomogeneous vacuum models}\label{sec:inhom}
The central assumption of BKL in the general \emph{inhomogeneous} context is
their locality conjecture.~\cite{lk63,bkl82} A physical justification of
asymptotic locality may heuristically be attempted in terms of the following
scenario: ultra strong gravity increasingly affects the causal structure as
the singularity is approached, and as a consequence particle horizons shrink
to zero size toward the singularity along each timeline. This prohibits
communication between different time lines in the asymptotic limit, and the
causal feature of asymptotically shrinking particle horizons along time lines
may hence be referred to as \textit{asymptotic silence}, while the associated
singularity is said to be \textit{asymptotically silent}.

In order to construct the solution in a sufficiently small spacetime
neighborhood of a generic spacelike singularity, Uggla {\it et
al}~\cite{uggetal03,rohugg05,heietal09} attempted to: \\
(i) capture asymptotic
silence and locality in a rigorous manner, and\\
(ii) contextualize the results
obtained in SH cosmology in terms of a general state space picture.

The tool needed
to satisfy (i) and (ii) while pursuing the goal of constructing an asymptotic
solution in a sufficiently small spacetime neighborhood of a generic
spacelike singularity in Refs.~\citen{uggetal03,rohugg05,heietal09} is a
reformulation of Einstein's field equations according to the following
prescription. Assume that a small neighborhood near the singularity can be
foliated with a family of spacelike surfaces such that the singularity
`occurs' simultaneously. Then `factor out' the expansion $\theta$ of the
normal congruence to the assumed foliation from Einstein's field equations by
first performing a conformal transformation (thereby respecting the expected
key causal structure),
\begin{equation}
{\bf g} = H^{-2}\,{\bf G},
\end{equation}
where $H$ is the Hubble variable associated with the expansion
($H=\sfrac{1}{3}\,\theta$) and ${\bf g}$ is the physical metric; since {\bf
g} has dimensional weight [length]$^{2}$ and $H$ [length]$^{-1}$ it follows
that the unphysical metric ${\bf G}$ is dimensionless. As the next step,
introduce an orthonormal frame for ${\bf G}$, or equivalently, a
corresponding conformal orthonormal frame for ${\bf g}$, where the unit normal of
the reference foliation is chosen to be the timelike vector field of the
orthonormal frame, and set the shift vector to zero so that the time lines are
tangential to that vector field. Finally, calculate Einstein's vacuum field
equations in terms of $H$ and the dimensionless frame and commutator
functions (or, equivalently, the connection) associated with the orthonormal
frame of ${\bf G}$. The conformal Hubble variable turns out to be minus the
deceleration parameter $q$ of the physical spacetime; moreover, it follows
from dimensional reasons that $q$ is algebraically determined by the
Raychaudhuri equation in terms of the other Hubble-normalized dimensionless
variables.

In the special case of spatial homogeneity, and a frame that is chosen so
that the spatial frame is tangential to the symmetry surfaces, the evolution
equations for the spatial Hubble-normalized frame variables (which
essentially are the spatial metric variables associated with ${\bf G}$)
decouple from the remaining equations. This follows from the fact that the
spatial frame derivatives of $H$ and the conformally Hubble-normalized
commutator functions are zero, since these quantities only depend on time due
to the symmetry assumption. Moreover, the evolution equation for $H$ also
decouples from the remaining equations since $H$ is the only variable that
carries dimension. This leaves a system of ODEs that coincides with the usual
Hubble-normalized dynamical system discussed in the previous section (when
restricted to the diagonalized class A vacuum case, otherwise it gives the
general Hubble-normalized SH equations), i.e., the above procedure provides a
general geometric setting for producing the usual Hubble-normalized dynamical
system that has been so successful in SH cosmology.

More importantly, however, is the following feature: the system of PDEs for
the general inhomogeneous case admits an invariant unphysical boundary subset
that is obtained by setting all spatial frame variables associated with ${\bf
G}$ to zero. Since this leads to that all spatial frame derivatives are set
to zero, this results in that the equations on this boundary subset form a
set of ODEs that coincides with the decoupled equation for $H$ and the
Hubble-normalized dynamical system of the SH case. The difference is that now
constants of integrations are only temporal constants, i.e., they now depend
on the spatial coordinates. Since the original expectation was that BKL
locality is due to asymptotic silence, this boundary was originally called
the silent boundary. Presumably asymptotic silence is a necessary condition
for BKL locality, but it turns out that asymptotic silence also admits other
possibilities, and for this reason the name of the invariant boundary subset
has been changed from the silent boundary to \emph{`the local
boundary'}~\cite{heietal12}.

It follows that the BKL assumption of locality, i.e., that the dynamics of an
individual timeline is asymptotically described by a SH model, therefore
corresponds to the statement that the asymptotic past dynamical evolution is
described by the local boundary, i.e., the BKL scenario obtains a precise
state space setting in the conformal Hubble-normalized approach. Since BKL is
about generic behavior, it is the past attractor on the local boundary that
is of interest, i.e., the asymptotic evolution along a generic timeline
should be described by the attractor for generic SH cosmology. The most
general of the SH models are those of Bianchi types IX, VIII and VI$_{-1/9}$.
The expected attractor for these models resides on a subset that consists of
the union of the vacuum Bianchi type I (Kasner) and II subsets. This,
together with setting the Hubble-normalized spatial frame variables to zero,
yields the same set of ODEs as those of the dimensionless coupled system in
the SH case. An advantage of the Hubble-normalized variables is that the
dimensionless variables are bounded on the expected past attractor subset,
i.e., even though the expected singularity is a scalar curvature singularity
with associated blow ups, the conformal Hubble-normalization leads to an
\emph{asymptotically regularized and bounded dimensionless state space}.
Furthermore, if one uses $H^{-1}$ as the dimensional variable, $H^{-1}$ tends
to zero towards the singularity, where $H^{-1} =0$ is an invariant
subset, associated with the fact that the singularity is a crushing singularity.

The above approach and picture was used in Ref.~\citen{heietal09} in order to
establish the consistency of the BKL scenario as well as the cosmological
billiards discussed by Damour and coworkers~\cite{dametal03} in the general
inhomogeneous context. However, the consistency of BKL locality and
cosmological billiards does not exclude other types of behavior. To gain
further insights about general oscillatory singularities in inhomogeneous
spacetimes it is natural to restrict investigations to models with two
commuting spacelike Killing vectors, so-called $G_2$ models. This was done in
Ref.~\citen{andetal05} where heuristical and numerical support was gained for
the BKL scenario in the Hubble-normalized state space context for an open set
of time lines, but this work also yielded evidence for `\emph{recurring
oscillatory spike formation}' for time lines forming 2D spatial surfaces, and
for time lines in their neighborhoods, for which it is \emph{not} possible to
neglect the Hubble-normalized spatial frame derivatives asymptotically, i.e.,
`asymptotic locality' is broken.

Oscillatory BKL behavior arises because certain `\emph{trigger}' variables
destabilizes the Kasner circle on the local boundary, which leads to
transitions between different fixed points on $\mathrm{K}^\mathrm{O}$ on the
local boundary. If such a variable goes through zero at a spatial coordinate,
i.e., at a spatial surface, then the `BKL transition' on the local boundary
cannot take place. Instead spatial derivatives grow and give rise to
\emph{spike transitions} between two different points on
$\mathrm{K}^\mathrm{O}$ on the local boundary. This had already been noticed
in special $G_2$ models such as the so-called Gowdy $T^3$ models, for reviews
and references, see Refs.~\citen{rin10} and~\citen{heietal12}, but the
asymptotic results of this phenomenon are quite different in such special
models, where the end result is that the evolution point wise approaches
$\mathrm{K}^\mathrm{O}$ on the local boundary in a non-uniform way, leading
to `permanent spikes'. This is \emph{not} what happens in the general case.
Instead infinitely oscillating recurring and transient spike formation,
leading to spike transitions from one Kasner state to another, take place,
which results in quite different non-uniform features.

The scenario in Ref.~\citen{andetal05} was also supported in
Ref.~\citen{lim04}, but more importantly, based on earlier work by Rendall
and Weaver~\cite{renwea01}\,, Lim managed to produce explicit
\emph{inhomogeneous} $G_2$ solutions by means of a solution generating
algorithm~\cite{lim08} that analytically describes spike transitions to high
accuracy~\cite{limetal09}, indeed the explicit solutions were essential in
order to accurately describe numerically several successive spike transitions
since this is quite challenging numerically. This work in turn led to
Ref.~\citen{heietal12} which yielded further analytic insights and numerical
progress concerning BKL and spike oscillations, as well as providing a
context for previous work on non-oscillatory models such as the $T^3$ Gowdy
models. It is this work we take as starting point for the discussion below
(for further details, see Ref.~\citen{heietal12}).

The description of $G_2$ models as well as the special $G_2$ spike solutions
naturally leads to the use of so-called Iwasawa frames. Since we choose the
reference time lines to be orthogonal to the assumed foliation, the shift
vector is set to zero which allows the line element to be written in
the form:
\begin{equation}\label{metrI}
ds^2 = - N^2(dx^0)^2 + \sum_{\alpha=1}^3 \exp(-2 b^{\alpha})\,{\bom}^\alpha\otimes {\bom}^\alpha,
\quad {\bom}^\alpha = \N{{\alpha}}{i}\,dx^i,
\end{equation}
where $i=1,2,3$ is summed over. In the general case without symmetries, $N=
N(x^\mu)$, $b^{\alpha} = b^{\alpha}(x^\mu)$, $\N{{\alpha}}{i} =
\N{{\alpha}}{i}(x^\mu)$ ($\mu=0,1,2,3$), and where an Iwasawa parametrization
of the spatial metric is implemented by setting
\begin{equation}\label{triang}
\Big(\,\N{{\alpha}}{i}\,\Big)_{\alpha, i} =
\begin{pmatrix}
1 & n_1 & n_2 \\
0 & 1 & n_3 \\
0 & 0 & 1
\end{pmatrix},
\end{equation}
where $n_1(x^\mu), n_2(x^\mu), n_3(x^\mu)$ (in the case of the $G_2$ models
with two commuting Killing vector fields $\partial_{x^1}$ and
$\partial_{x^2}$, the use of a symmetry adapted frame leads to that all
variables depend on $x^0$ and $x^3$ only).

\begin{figure}[ht]
\centering
\includegraphics[width=0.30\textwidth]{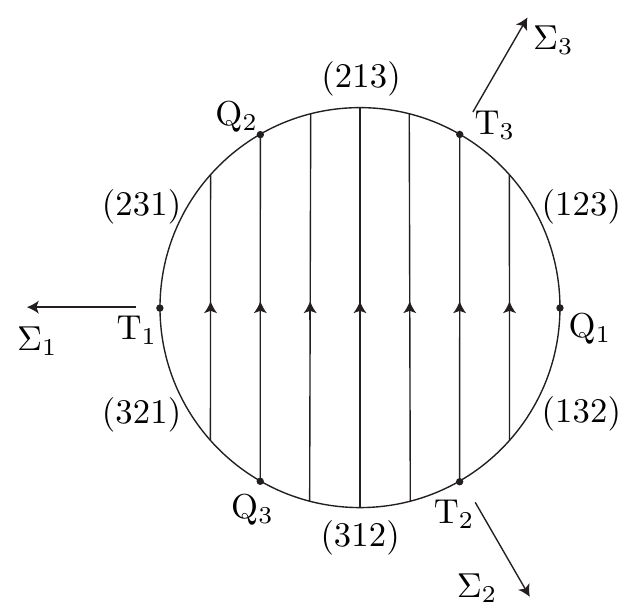}\qquad
\includegraphics[width=0.30\textwidth]{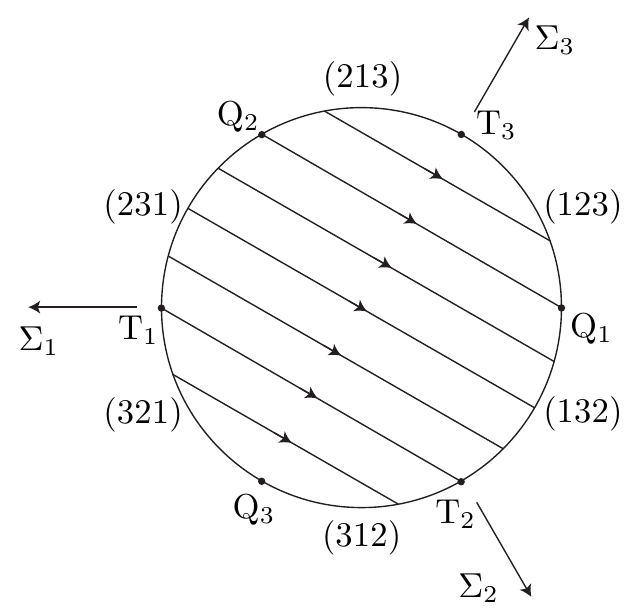}
        \caption{Projections  onto $(\Sigma_1, \Sigma_2, \Sigma_3)$ space
                 of the two types of frame transitions,
        $\mathcal{T}_{R_1}$ and $\mathcal{T}_{R_3}$,
          that exist in the $G_2$ case.
          As usual the direction of time indicated by the arrows
          is towards the past.}
           \label{frametrans}
\end{figure}

It follows that the diagonal conformally Hubble-normalized shear variables,
$\Sigma_{\alpha}\equiv \Sigma_{\alpha\alpha}  $, of the timelike reference
congruence are given by:
\begin{equation}
\Sigma_\alpha :=  - (1 +  (HN)^{-1}\partial_{x^0}b^\alpha),
\qquad\qquad \Sigma_1 + \Sigma_2 + \Sigma_3 = 0.
\end{equation}
Other central variables are
\begin{subequations}\label{ein.3}
\begin{align}
N_1 &:= \sfrac{1}{2}H^{-1}\tilde{N}_1\,\exp(b_2 + b_3 - b_1),\\
R_{1} &:= -\sfrac{1}{2}\exp(b^3 - b^2)\,(HN)^{-1}\partial_{x^0}(n_3),\\
R_{3} &:= -\sfrac{1}{2}\exp(b^2 - b^1)\,(HN)^{-1}\partial_{x^0}(n_1),
\end{align}
\end{subequations}
where $N_1$ is a Hubble-normalized spatial commutator function ($\tilde{N}_1$
is a rather complicated expression that involves spatial derivatives
of $n_1, n_2, n_3$, see Ref.~\citen{heietal09}), $R_1 = \Sigma_{23}$, $R_3 =
\Sigma_{12}$, while $R_2=-\Sigma_{31}$ can be set to zero for the vacuum
$G_2$ models. The Kasner circle $\mathrm{K}^\mathrm{O}$ on the local boundary
is described in the same way as in the spatially homogeneous case, so that
 the only non-zero variables on that circle are the temporally constant, but spatially
dependent, Hubble-normalized diagonal shear variables; define the total shear quantity
$\Sigma^2 := \frac16(\Sigma_1^2 + \Sigma_2^2 + \Sigma_3^2)$. The variables
$N_1$, $R_1$ and $R_3$ are `trigger variables' that destabilize
$\mathrm{K}^\mathrm{O}$ leading to transitions between different fixed points
on $\mathrm{K}^\mathrm{O}$; $N_1$ yields $\mathcal{T}_{N_1}$ Bianchi type II
transitions on the local boundary (see Fig.~\ref{Fig:alltypeII}) while $R_1$
and $R_3$ result in so-called $\mathcal{T}_{R_1}$ and $\mathcal{T}_{R_3}$
`frame transitions' (trajectories that form straight lines in $(\Sigma_1,
\Sigma_2, \Sigma_3)$ space with $\Sigma_1 = \mathrm{const}$ and $\Sigma_3 =
\mathrm{const}$, respectively), i.e., rotations of the spatial frame that
transfer one Fermi propagated (i.e., gyroscopically fixed) Kasner state on
$\mathrm{K}^\mathrm{O}$, i.e., a fixed point, to another, keeping the gauge
invariant (spatially dependent) Kasner parameter $u$ fixed, see
Fig.~\ref{frametrans}.

In the general case one of the trigger variables might become zero at a value
of a spatial coordinate. In this case the role of the trigger is replaced by
increasing spatial gradients which act as a `spike trigger' that changes the
dynamical state. If $R_1$ or $R_3$ goes through zero this asymptotically
leads to so-called false spikes, which are gauge features which we will
refrain from discussing, while if $N_1$ goes through zero this results in
recurring `true spikes.'

\begin{figure}[ht]
 \centering{
  \includegraphics[height=0.45\textwidth]{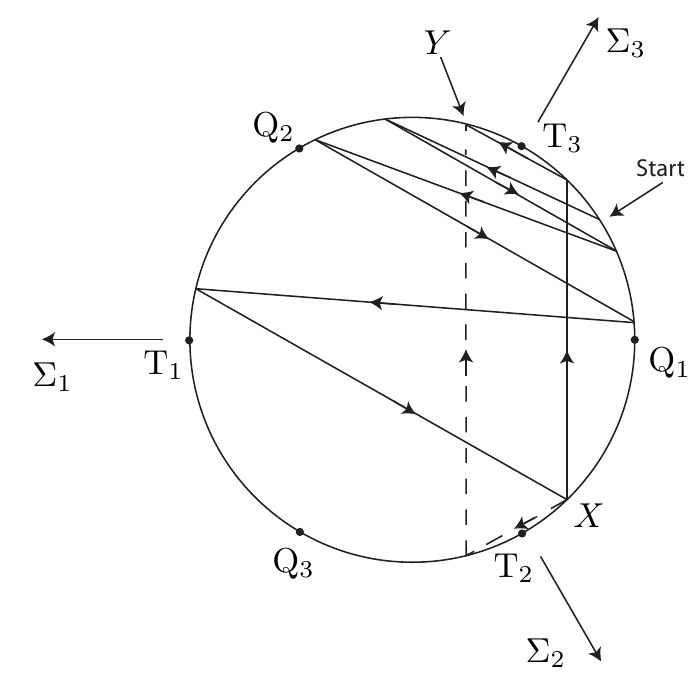}}
\caption{An Iwasawa chain is an (in general) infinite heteroclinic chain on the local boundary
consisting of $\mathcal{T}_{N_1}$, $\mathcal{T}_{R_1}$ and
$\mathcal{T}_{R_3}$ transitions.
The paths are not unique; e.g., at the point ${X}$ two continuations are possible,
a $\mathcal{T}_{R_1}$ or a $\mathcal{T}_{N_1}$ transition (dashed line).
However, the different paths rejoin at the point ${Y}$ where
the chains continue with a (not shown) $\mathcal{T}_{R_3}$ transition.}
\label{BKLSequence}
\end{figure}

In the case the dynamics along a timeline in a general $G_2$ model approaches
the local boundary in the state space and $N_1R_1R_3 \neq 0$, the dynamics is
BKL-like and can be described asymptotically by concatenation of
$\mathcal{T}_{N_1}$, $\mathcal{T}_{R_1}$ and $\mathcal{T}_{R_3}$ transitions
into heteroclinic (BKL) `Iwasawa chains' on the local boundary (one for each
spatial point), see Fig~\ref{BKLSequence}.\footnote{In contrast to Bianchi
types IX and VIII there exist several triggers in sectors $(132)$ and $(321)$
on $\mathrm{K}^\mathrm{O}$. These might be simultaneously activated, giving
rise to a `multiple' transition that results in a final state on
$\mathrm{K}^\mathrm{O}$ that coincides with the fixed point obtained by
successively applying the present `single' frame transitions that constitute
the boundary of the multiple transition. However, as discussed in
Ref.~\citen{heietal09}, multiple transitions are not expected to play a
generic role in the asymptotic dynamics of solutions towards a generic spacelike
singularity, and we therefore only discuss the `single' transitions
$\mathcal{T}_{N_1}$, $\mathcal{T}_{R_1}$ and $\mathcal{T}_{R_3}$.}

\begin{figure}[ht]
\centering
          \includegraphics[height=0.37\textwidth]{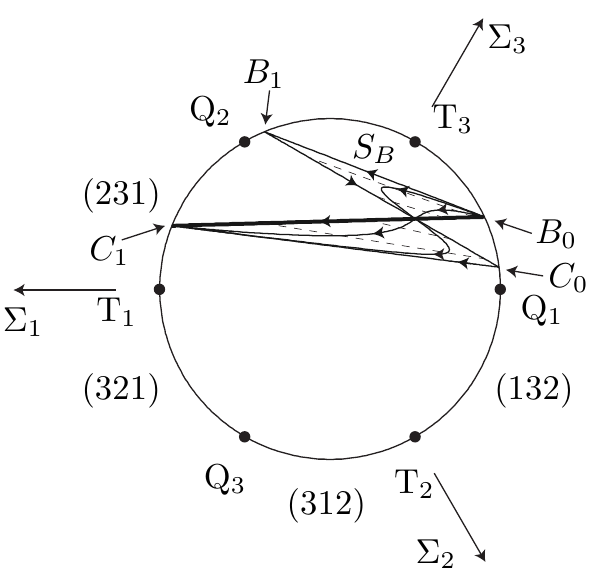}\qquad
          \includegraphics[height=0.37\textwidth]{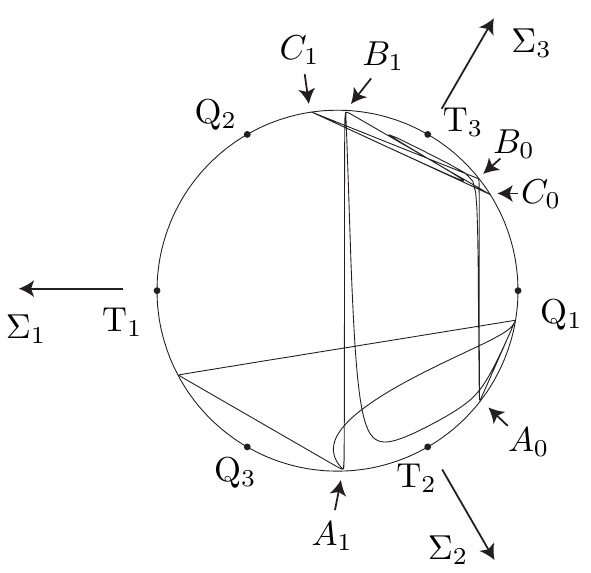}
          \caption{A high velocity spike transition $\Thi$ (figure to the left) and a so-called
          joint low/high velocity spike transition $\Tco$ (figure to the right),
          the basic building blocks for spike concatenation, projected onto
          $(\Sigma_1, \Sigma_2, \Sigma_3)$ space. Each $\Thi$/$\Tco$ corresponds to a family of curves
          parameterized by the spatial coordinate $x^3$. The thick (straight) line in the $\Thi$ case
          is the trajectory that describes the evolution of the time lines of the
          spike surface $x^3 = 0$, a so-called spike surface trajectory.
          The thin curves describe the trajectories of time lines $0 \neq |x^3| \not\gg 1$. The
          thin straight lines describe evolution along $|x^3| \gg 1$ time lines; these are short (BKL) Iwasawa chains,
          i.e., sequences of Bianchi type II and frame transitions on the local boundary.
         For further details and explanations, see Ref.~\citen{heietal12}.}
        \label{G2hlconc2}
\end{figure}

However, if $N_1=0$ at a value of $x^3$ this gives rise to so-called
\emph{spike chains}. In contrast to the BKL scenario, if $N_1$ goes through
zero it affects a whole family of time lines whose dynamics differ, which is
illustrated by a family of curves in the state space picture, conveniently
projected onto $(\Sigma_1,\Sigma_2,\Sigma_3)$ space, in Fig.~\ref{G2hlconc2}.
Nevertheless, the trajectories of \emph{all} `spike' time lines rejoin at a
common fixed point on $\mathrm{K}^\mathrm{O}$, and it is therefore these
instances that define the natural `concatenation blocks'. These turn out to
consist of either so-called `high-velocity' solutions, which hence are
referred to as high-velocity transitions $\Thi$, and so-called low-velocity
solutions combined with $\mathcal{T}_{R_1}$ transitions and part of
high-velocity solutions that form a \textit{joint low/high velocity spike
transition}, $\Tco$ (Fig.~\ref{G2hlconc2}). In addition, there are so-called
Bianchi type II spiky features, which we will not discuss, but instead refer
to Ref.~\citen{heietal12}.

The effect of $\Thi$ as well as $\Tco$ for a family of time lines is to
transform a Kasner state described by the Kasner parameter $u^{\mathrm{i}}$
to another Kasner state $u^{\mathrm{f}}$ (the time direction is towards the
singularity) according to a map that is obtained by applying~\eqref{BKLMap}
twice, which results in~\cite{heietal12}
\begin{equation}\label{spikemap}
u^{\mathrm{f}} =
\begin{cases}
u^{\mathrm{i}} - 2 & u^{\mathrm{i}} \in [3 ,\infty)\,, \\
(u^{\mathrm{i}} - 2)^{-1} & u^{\mathrm{i}} \in [2,3] \,,\\
\big((u^{\mathrm{i}} - 1)^{-1} - 1 \big)^{-1} & u^{\mathrm{i}} \in [ 3/2 ,2] \,,\\
(u^{\mathrm{i}} - 1)^{-1} - 1 & u^{\mathrm{i}} \in [1, 3/2]\:.
\end{cases}
\end{equation}

\begin{figure}[ht]
\centering
        \includegraphics[height=0.36\textwidth]{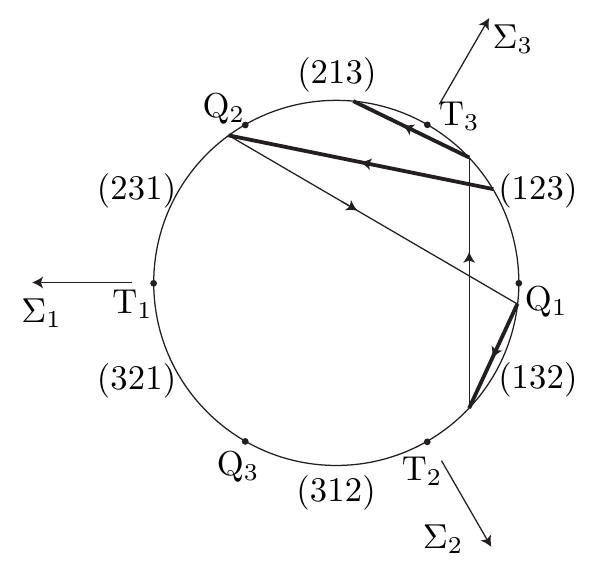}
        \caption{An example of (parts of) a spike chain with
        $\Thi$--$\mathcal{T}_{R_3}$--$\Tco$  where only spike
        surface trajectories are shown (i.e., the picture
        depicts the evolution along time lines that are located at the spike
        surface). At the spike surface $\Tco$ consists of joining a so-called
        low velocity solution (here represented by the thick line near sector $(132)$),
        a $\mathcal{T}_{R_1}$ frame transition, and a $\Thi$ transition
        with each other, to form one concatenation block, for which, by definition
        \emph{all} time lines begin and end at the same two distinct fixed points
        at the Kasner circle (a feature that is not depicted in the figure).}
        \label{G2hlconc}
\end{figure}

Iteration of spike concatenation blocks lead to oscillations of different
Kasner states, common to all time lines affected by the recurring spike and
frame transitions, see Fig.~\ref{G2hlconc} for concatenation of spike surface
trajectories. As shown and estimated in Ref.~\citen{heietal12}, the spatial
size of the spike shrinks towards the singularity and asymptotically this is
expected to lead to a new type of non-uniformity. The dynamics along the
time lines that form the spike surface, at the value of $x^3$ for which
$N_1(x^3)=0$, is given by the dynamics at that $x^3$ of an invariant subset
that consists of the exact inhomogeneous $G_2$ solutions found by Lim (the
so-called high and low velocity spike solutions) and the Kasner subset on the
local boundary, while the dynamics along the surrounding time lines is
described by the past (billiard) attractor (see Ref.~\citen{heietal09}) on
the local boundary (a completely different part of the state space than the
`Lim solution subset'). This is in stark contrast to the asymptotic
non-uniformities found in special $G_2$ models for which the `trigger
variable' $R_1$ is identically zero as occurs for the $T^3$ Gowdy models. There
non-uniformities are associated with `permanent' spikes that occur because
the natural concatenation block $\Tco$ is `interrupted' in its evolution,
thereby being `cut in half,' due to the absence of $R_1$, which leads to a
picture that is quite misleading regarding generic spacelike singularities,
see Ref.~\citen{heietal12} for details.

%

\section{Discussion}\label{sec:disc}

The concatenation blocks of `BKL' chains and spike chains are all closely
related in a hierarchical manner. Firstly, the Kasner circle
$\mathrm{K}^\mathrm{O}$ on the local boundary is the simplest and most basic
common ingredient. Secondly, BKL chains consisting of frame and Bianchi type
II transitions on the local boundary describe the dynamical evolution of
spike transitions far from the spike surface of recurring spikes. Thirdly,
all solutions that form the concatenation blocks are related to each other in
a hierarchical manner via a solution generating
algorithm~\cite{renwea01,lim08}. The Kasner solutions in the form associated
with the Kasner circle $\mathrm{K}^\mathrm{O}$ in the spatially homogeneous
Bianchi type I model act as the initial seed solutions that yield the Kasner
solutions in a rotating frame which describe the Kasner frame transitions. That
form for the Kasner solutions is subsequently the seed for the Bianchi type
II vacuum solutions, which in turn act as the seed yielding a frame rotated version of the
Bianchi type II solutions, which form the so-called false spike solutions.
Using these solutions as the seed then results in the spike solutions.

The hierarchical structures discussed in the previous paragraph are not the
only ones; we have seen earlier that the Bianchi Lie contraction hierarchy
plays an essential role for the asymptotics of the Bianchi models, and that
this to a large extent determines the asymptotics on the local boundary. This
is in turn part of a greater hierarchy. The primary importance of the $G_2$
models regarding generic singularities is not the models themselves but the
following important property: in the $G_2$ models certain Hubble-normalized
spatial frame variables decouple due to the imposed symmetries. Setting these
to zero in the general case without symmetries yields an invariant boundary
subset, called the \emph{partially local $G_2$ boundary subset}, with
equations that are identical to the coupled set of equations describing the
$G_2$ models, but the solutions on this subset involve all the coordinates.
The situation is therefore completely analogous to the relationship between
the local boundary and the spatially homogeneous models.

The partially local $G_2$ boundary subset was previously referred to as a
partially silent boundary subset, since it is associated with some solutions
that have singularities that break asymptotic silence, which corresponds to
the existence of directions in which particle horizons extend to infinity,
see Ref.~\citen{limetal06}. However, since it is also associated with
recurring spikes, which, due to the prominence of oscillating Kasner states,
seem likely to be connected with asymptotic silence, it is  motivated to
refer to it as the partially local $G_2$ boundary subset, rather than a
partially silent boundary subset.

Going beyond the $G_2$ assumption and looking at models with fewer isometries
not only shifts attention to the partially local $G_2$ boundary, it
leads to potentially new phenomena as well. In the general case spike
surfaces are no longer planes and they can intersect in curves that in turn
can intersect at points, which lead to new spike dynamics. At present it is
not known whether such intersections persist or recur, although weak
numerical evidence suggests that intersections only occur momentarily. If
this is correct, it follows that spike intersections are irrelevant
asymptotically. This would then imply that the BKL picture in combination
with $G_2$ spike oscillations may capture the essential features of generic
spacelike singularities.\cite{heietal12}

The presently discussed recurring spikes are located at fixed spatial
locations due to the choice of initial data. In general recurring spikes are
moving in space. It may be that they asymptotically freeze, but this is an
open issue. If they do freeze, then our present knowledge about recurring
spikes represents the first step in understanding general recurring spike behavior,
otherwise perhaps not. Apart from this, there are many other unresolved
questions pertaining to recurring spike behavior and generic spacelike
singularities. For example, how many spikes can form? Can spikes annihilate?
Are there generic singularities without recurring spikes? Are there generic
singularities with a dense set of recurring spikes? Are there boundary
conditions associated with special physical conditions that explain the
existence of recurring spikes? Are oscillatory singularities not cosmological
in nature but instead the spacelike part of generic black hole singularities?
These issues, together with those mentioned for the spatially homogeneous
case, illustrate that we are only at the beginning of understanding generic
singularities, even though considerable progress has been accomplished during
the last few years.

\subsection*{Acknowledgements}
It is a pleasure to thank Mark Heinzle and Woei Chet Lim for joint work in
this area, and many helpful and stimulating discussions that have made this
article possible. I would also like to thank Bob Jantzen for help with the
manuscript and especially for his support and friendship.

\end{document}